\documentclass[conference]{IEEEtran}
\IEEEoverridecommandlockouts

\usepackage{cite}
\usepackage{amsmath,amssymb,amsfonts}
\usepackage{hyperref}
\usepackage[ruled,noend,linesnumbered]{algorithm2e}
\usepackage{listings}
\usepackage{graphicx}
\usepackage{textcomp}
\usepackage{slashbox}
\usepackage{xcolor}
\usepackage{bm}

\title{Reduced Kernel Dictionary Learning}

\author{
    \IEEEauthorblockN{Denis C. ILIE-ABLACHIM} \\
    \IEEEauthorblockA{
        \textit{Faculty of Automatic Control and Computers}\\
        \textit{University Politehnica of Bucharest}\\
        denis.ilie\_ablachim@upb.ro
    }
    \and
    \IEEEauthorblockN{Bogdan DUMITRESCU}\\
    \IEEEauthorblockA{
        \textit{Faculty of Automatic Control and Computers}\\
        \textit{University Politehnica of Bucharest}\\
        bogdan.dumitrescu@upb.ro
    }
    \thanks{This work was supported by grants of the Romanian Ministry of Research, Innovation and Digitization,
        CCCDI - UEFISCDI, project numbers PN-III-P2-2.1-PED-2019-3248
        and PN-III-P4-PCE-2021-0154, within PNCDI III.
    }
}

\begin{document}
\maketitle

\begin{abstract}
In this paper we present new algorithms for training reduced-size nonlinear
representations in the Kernel Dictionary Learning (KDL) problem.
Standard KDL has the drawback of a large size of the kernel matrix when the
data set is large. There are several ways of reducing the kernel size,
notably Nystr\"om sampling.
We propose here a method more in the spirit of dictionary learning, where
the kernel vectors are obtained with a trained sparse representation
of the input signals.
Moreover, we optimize directly the kernel vectors in the KDL process,
using gradient descent steps.
We show with three data sets that our algorithms are able to provide better
representations, despite using a small number of kernel vectors,
and also decrease the execution time with respect to KDL.
\end{abstract}

\section{Introduction}
Dictionary Learning (DL) is a representation learning method that aims to find a sparse representation for a set of signals, $\bm{Y}$, represented as a matrix with $N$ columns (signals) of size $m$. The representation is achieved by computing a dictionary $\bm{D}$ of size $m \times n$ and a sparse representation $\bm{X}$ of size $n \times N$ such that a good approximation $\bm{Y} \approx \bm{DX}$ is obtained. Most applications with dictionary learning are in image denoising, inpainting, signal reconstruction, clustering or classification.

In this paper we present new methods of dictionary learning that produce sparse representations in both linear and nonlinear spaces, starting from the Kernel Dictionary Learning (KDL) idea \cite{NPNC13,TRS14}.
In a first approach, this is done in two stages; a linear representation is built and the resulting optimized dictionary is used unchanged in the nonlinear space as kernel vectors during the training procedure.
In the second approach, the kernel vectors are optimized alongside with the nonlinear representation. The main advantage of this method is the use of a reduced matrix, $\bm{D}$, containing the kernel vectors, which is also the dictionary in a standard DL problem. By the use of the reduced matrix, the built-in kernel is smaller and thus the problem complexity is reduced.

As KDL has a high complexity when $N$ is large (and so the kernel matrix is large), solutions have been adopted from other problems where kernels appear. In Large-Scale Kernel Machines, various kernel enhancement or resizing strategies have been used, such as Nystr\"om Sampling \cite{golts2016linearized}, \cite{kumar2012sampling} or Random Fourier Features (RFF) \cite{rahimi2007random}. The first one computes a rank $\hat{m}$ approximation $\hat{\bm{K}}$ of the kernel matrix $\bm{K}$. The whole procedure consists in approximating the nonlinear mapping function $\varphi(\bm{Y})$ with a matrix $\hat{\bm{Y}}$, containing a compressed version of the original signals. Compared to the original problem, the new problem no longer requires a high computational cost. The Random Fourier Features method proposes to map the input signals to a randomized low-dimensional feature space. More exactly, the inner product, used in the kernel trick, is replaced with a randomized map $k(\bm{x}, \bm{y}) = \varphi(\bm{x})^{\top}\varphi(\bm{y}) = z(\bm{x})^{\top} z(\bm{y})$, where $z: \mathbb{R}^{m} \mapsto \mathbb{R}^{\hat{m}}$ and $\hat{m} \ll m$. Both methods enable the use of fast linear methods, which will further use the resulting features.
The use of reduced kernels dates from Support Vector Machine times, significant contributions being \cite{LeeMa2001rsvm,lin2003study}, with applications in classification \cite{fung2002minimal}.
Other learning methods where a reduced kernel appears can be found in \cite{deng2013reduced}.
In these early works, kernel vectors are usually selected (randomly or with some heuristic) as a subset of input signals.
Finally, there are KDL substitutes like \cite{HuTan2018nonlinear}, where the nonlinear transformation is performed by a neural encoder-decoder, with standard DL on the encoded signals.

The paper is organized as follows. In Section \ref{sec:KDL} we review the standard DL and KDL problems and the most common algorithm for solving them.
In Section \ref{sec:RKDL} we present our own contribution, named the Reduced Kernel Dictionary Learning (RKDL) problem, under three different scenarios, in which the kernel vectors are:
a) the result of a DL problem, solved a priori;
b) optimized together with the other KDL variables, using gradient descent,
with an objective that is directly related to that of the KDL problem;
c) optimized like before, but using a mixed objective that combines the
nonlinear representation of the signals with the linear representation
of the kernel vectors.
The algorithms for b) and c) are new and their representation error is smaller.
Section \ref{sec:Experiments} contains the experimental results, obtained on three public data sets: Digits, MNIST \cite{lecun1998gradient} and CIFAR-10 \cite{krizhevsky2009learning}, under the three proposed scenarios.

\section{Kernel Dictionary Learning}
\label{sec:KDL}

The DL problem is formulated as following
\begin{equation}
\begin{array}{ll}
\displaystyle\min_{\bm{D}, \bm{X}} & \|\bm{Y}-\bm{D} \bm{X}\|_{F}^{2} \\
\text { s.t. } & \left\|\bm{x}_{\ell}\right\|_{0} \leq s_{x}, \ell=1:N \\
& \left\|\bm{d}_{j}\right\|=1, j=1:n,
\end{array}
\label{eq:DL}
\end{equation}
where $\left\|\cdot\right\|_{0}$ represents the $0$-pseudo-norm, $s_{x}$ is the sparsity level and $\bm{d}_{j}$ is a column (named also atom) of $\bm{D}$.

The standard dictionary learning problem can be solved by using simple strategies. In order to overcome the nonconvexity and the huge dimension of the problem, the most usual optimization procedure iterates two basic steps and is also known as DL by Alternate Optimization. In this way, the problem is divided in two subproblems: sparse coding and dictionary update. By alternating these two stages for a given number of iterations, good local solutions can be obtained. A simple iteration consists of computing the sparse representation $\bm{X}$, using Orthogonal Matching Pursuit (OMP) \cite{omp} while the dictionary $\bm{D}$ is fixed, and then updating the dictionary atoms successively while the sparse representation is fixed.
There are several methods of dictionary learning \cite{dlaa}, but the one of interest to us is AK-SVD \cite{k-svd}, \cite{rubinstein2008efficient} due to its low complexity and good performance.

In the DL problem, the input data are modeled through a linear representation, which in some cases may be seen as a limitation.
In order to overcome this drawback, kernel representations can be used for a better quantification of similarities or differences between input vectors.

The kernel representation is an extension to nonlinearity.
We do this by associating to a data vector $\bm{y} \in \mathbb{R}^{m}$ a feature vector $\varphi(\bm{y}) \in \mathbb{R}^{\tilde{m}}$, where $\varphi : \mathbb{R}^{m} \rightarrow \mathbb{R}^{\tilde{m}}$ is a nonlinear function.
Typically, Mercer kernels are used, which can be expressed as a scalar product of feature vector functions $k(\bm{x}, \bm{y})={\varphi}(\bm{x})^{\top} {\varphi}(\bm{y})$. In the last form, the scalar product can be replaced with a specific function definition, such as radial basis function (RBF) $k(\bm{x}, \bm{y}) = \exp(-\frac{\|\bm{x} - \bm{y}\|^{2}_{2}}{2\sigma^{2}})$ or polynomial kernel $k(\bm{x}, \bm{y}) = (\bm{x}^{\top}\bm{y} + \alpha)^{\beta}$.

The Kernel Dictionary Learning (KDL) \cite{NPNC13}, \cite{TRS14} problem is
\begin{equation}
\begin{array}{ll}
\displaystyle\min_{\bm{A}, \bm{Z}} & \|\varphi(\bm{Y})-\varphi(\bm{Y})\bm{A} \bm{Z}\|_{F}^{2} \\
\text { s.t. } & \left\|\bm{z}_{\ell}\right\|_{0} \leq s_{z}, \ell=1:N \\
& \left\|\varphi(\bm{Y})\bm{a}_{j}\right\|=1, j=1:n,
\end{array}
\label{eq:KDL}
\end{equation}
where $\varphi(\bm{Y})$ represents the nonlinear extension of data  and $\varphi(\bm{Y})\bm{A}$ is the kernel dictionary, where $\bm{A}$ is the coefficients matrix of size $N \times p$. Depending on the used data set, the problem can be difficult to solve due the large kernel matrix $\bm{K}_{YY} = \varphi(\bm{Y})^{\top}\varphi(\bm{Y})$ that results from the trace form of the objective function. In this case, the problems with large data sets can involve large volume of memory and long execution times.
The KDL problem can also be solved by alternate optimization.
The sparse representation is computed according to the Kernel OMP algorithm \cite{NPNC13}.
The columns of the matrix $\bm{A}$ are sequentially updated with an algorithm
inspired by AK-SVD, while the representation matrix $\bm{X}$ is fixed.

Both algorithms, AK-SVD and Kernel AK-SVD, alongside with the sparse representation computation are presented in \cite{dlaa}.

\section{Reduced Kernel Dictionary Learning}
\label{sec:RKDL}

Nonlinear space can extend the horizon of data representation.
However, KDL has disadvantages when the number of available signals is large. The size of the kernel increases in proportion to the size of the data. Thus, the problem becomes more complex from a numerical point of view. 

In order to overcome this limitations we use a reduced space on which the kernel matrix is built. This strategy is implemented by using as kernel vectors not the full set of
signals, $\bm{Y}$, but a smaller set of vectors, $\bm{D}$,
trained with DL as a dictionary for linear representations,
thus replacing \eqref{eq:KDL} with
\begin{equation}
\begin{array}{ll}
\displaystyle\min_{\bm{A}, \bm{Z}} & \|\varphi(\bm{Y})-\varphi(\bm{D})\bm{A} \bm{Z}\|_{F}^{2} \\
\text { s.t. } & \left\|\bm{z}_{\ell}\right\|_{0} \leq s_{z}, \ell=1:N \\
& \left\|\varphi(\bm{D})\bm{a}_{j}\right\|=1, j=1:n.
\end{array}
\label{eq:RKDL}
\end{equation}

This problem has two advantages that can be used as needed. If the number of training signals is very large, a dictionary with a much smaller number of atoms can be used. On the other hand, we have problems where the number of training signals is small or the given signals are not representative enough for the representation problem. In this case it is recommended to use large dictionaries. In our work the case of interest is the first one.

\subsection{Standard Reduced Kernel Dictionary Learning}

A first approach to \eqref{eq:RKDL} consists of solving it in two steps.
In the first step we design $\bm{D}$ by solving a DL problem and thus
obtaining a trained dictionary.
We call this method RKDL-D, due the use of matrix $\bm{D}$; it was introduced in \cite{RKDL}.
Here is a brief reminder of the atom update step of RKDL-D.
(Sparse representation can be easily derived from Kernel OMP.)

Expressing the objective of \eqref{eq:RKDL} in its trace form
and isolating the current atom $a_{j}$, we can write
\[
\begin{split}
\operatorname{Tr}\left[\left(\varphi^{\top}(\bm{Y})-\sum_{i \neq j} \bm{z}_{i} \bm{a}_{i}^{\top} \varphi^{\top}(\bm{D})- \bm{z}_{j} \bm{a}_{j}^{\top} \varphi^{\top}(\bm{D})\right)\right. \\ \left. \left(\varphi(\bm{Y})-\varphi(\bm{D}) \sum_{i \neq j} \bm{a}_{i} \bm{z}_{i}^{\top}-\varphi(\bm{D}) \bm{a}_{j}\bm{z}_{j}^{\top}\right) \right].
\end{split}
\]
We compute the partial derivatives with respect to the current atom $\bm{a}_{j}$
\[
\frac{\partial \operatorname{Tr}(\cdot)}{\partial \bm{a}_{j}} = 2\|\bm{z}_{j}\|^{2} \bm{K}_{DD} \bm{a}_{j}+2 \bm{K}_{DD} \bm{R} \bm{z}_{j} - 2 \bm{K}_{YD} \bm{z}_{j} \\
\]
and the current sparse representation vector $\bm{z_{j}}$
\[
\frac{\partial \operatorname{Tr}(\cdot)}{\partial \bm{z}_{j}} = 2 \bm{a}_{j}^{\top} \bm{K}_{DD} \bm{a}_{j} \bm{z}_{j} + 2 \bm{R}^{\top} \bm{K}_{DD} \bm{a}_{j} - 2 \bm{K}_{YD} \bm{a}_{j}.
\]
where we have used the notations $\bm{K}_{YD} = K(\bm{Y}, \bm{D}) = \varphi(\bm{Y})^{\top}\varphi(\bm{D})$, $\bm{K}_{DD} = K(\bm{D}, \bm{D})$ and $\displaystyle\bm{R} = \sum_{i \neq j} \bm{a}_{i} \bm{z}_{i}^{\top}$.

Setting these derivatives to zero, we obtain the optimal atom
(for fixed representation)
\begin{equation}
\bm{a}_{j}= \left(\|\bm{z}_{j}\|_{2}^{2} \bm{K}_{DD} \right)^{-1} (\bm{K}_{YD} + \bm{K}_{DD} \bm{R}) \bm{z}_{j}
\end{equation}
and optimal representation (for fixed atom)
\begin{equation}
\bm{z}_{j} = \left( \bm{K}_{YD} - \bm{R}^{\top} \bm{K}_{DD} \right) \bm{a}_{j}.
\end{equation}

The RKDL-D procedure of updating atoms is summarized in Algorithm \ref{alg:RKDL}.
For brevity, we did not use special notations, but only the signals from
$\bm{Y}$ where $\bm{a}_j$ appears in the representation are involved in
the computation.

\begin{algorithm}[ht!]
\KwData{complementary kernel matrix $\bm{K}_{DD} \in \mathbb{R}^{p \times p}$ \\ \hspace{0.85cm}
partial kernel matrix $\bm{K}_{YD} \in \mathbb{R}^{N \times p}$ \\ \hspace{0.85cm}
current kernel dictionary $\bm{A} \in \mathbb{R}^{N \times n}$ \\ \hspace{0.85cm}
representation matrix $\bm{Z} \in \mathbb{R}^{n \times N}$}
\KwResult{updated kernel dictionary $\bm{A}$, representation $\bm{Z}$}
Compute sum $\displaystyle\bm{S}=\sum_{i=1}^{n}\bm{Z}_{i}^{\top}\bm{a}_{i}^{\top}$ \\
\For{$j=1$ {\bf to} $n$}{
Modify sum: $\bm{R}=\bm{S} - \bm{Z}_{j} \bm{a}_{j}^{\top}$  \\
Update atom: $\bm{a}_{j}= \left(\|\bm{z}\|_{2}^{2} \bm{K}_{DD} \right)^{-1} (\bm{K}_{YD}^{\top} + \bm{K}_{DD}\bm{R}) \bm{Z}_{j}$ \\
Normalize atom: $\bm{a}_{j} \leftarrow \bm{a}_{j} / \left(\bm{a}_{j}^{\top} \bm{K}_{DD} \bm{a}_{j}\right)^{\frac{1}{2}}$ \\ 
Update representation: $\bm{Z}_{j}^{\top} \leftarrow (\bm{K}_{YD} - \bm{R} \bm{K}_{DD}) \bm{a}_j$ \\
Recompute error: $\bm{S} = \bm{R} + \bm{Z}_{j} \bm{a}_{j}^{\top}$ \\
}
\caption{RKDL-D -- update step}
\label{alg:RKDL}
\end{algorithm}

\subsection{Optimized Reduced Kernel Dictionary Learning}

The improvement that we propose here is to update the dictionary $\bm{D}$,
containing the kernel vectors,
during the nonlinear optimization procedure.
We keep the idea of alternate optimization.
The matrices $\bm{Z}$, $\bm{A}$ and $\bm{D}$ are updated successively.
As above, for $\bm{Z}$ we use Kernel OMP and the atoms of 
$\bm{A}$ are updated as described by Algorithm \ref{alg:RKDL}.
Updating the dictionary $\bm{D}$ must be done with a different procedure, detailed below.
Since the dictionary $\bm{D}$ is updated together with the nonlinear representation, we call the resulting method Optimized Reduced Kernel Dictionary Learning (ORKDL-D).

In order to solve the optimization problem for $\bm{D}$, we update each column $\bm{d_{j}}$ independently, by the use of the trace form of the objective function of \eqref{eq:RKDL}:
\[
\operatorname{Tr}\left[\bm{K}_{YY} - 2 \bm{K}_{YD} \bm{A} \bm{Z} +  \bm{Z}^{\top} \bm{A}^{\top} \bm{K}_{DD} \bm{A} \bm{Z} \right].
\]
We compute the partial derivatives with respect to the $i$th element of the current column $\bm{d}_{j}$, for both nonlinear terms
\[
\arraycolsep 1pt
\begin{array}{rcl}
\frac{\partial \operatorname{Tr}\left[\bm{K}_{YD} \bm{A} \bm{Z} \right]}{\partial \bm{d_{ij}}} & = & \operatorname{Tr} \left[ \left( \frac{\partial \operatorname{Tr} \left[ \bm{K}_{YD} \bm{A} \bm{Z}  \right]}{\partial \bm{K}_{YD}} \right)^{\top} \cdot \frac{\partial \bm{K}_{YD}}{\partial \bm{d_{ij}}} \right] \\ & = & \operatorname{Tr} \left[ \bm{A} \bm{Z} \cdot \frac{\partial \bm{K}_{YD}}{\partial \bm{d_{ij}}} \right]
\end{array}
\]
and
\[
\arraycolsep 1pt
\begin{array}{rcl}
\frac{\partial \operatorname{Tr}\left[\bm{Z}^{\top} \bm{A}^{\top} \bm{K}_{DD} \bm{A} \bm{Z} \right]}{\partial \bm{d_{ij}}} & = & \operatorname{Tr} \left[ \left( \frac{\partial \operatorname{Tr} \left[ \bm{Z}^{\top} \bm{A}^{\top} \bm{K}_{DD} \bm{A} \bm{Z}  \right]}{\partial \bm{K}_{DD}} \right)^{\top} \cdot \frac{\partial \bm{K}_{DD}}{\partial \bm{d_{ij}}} \right] \\
& = & \operatorname{Tr} \left[ \bm{A} \bm{Z} \bm{Z}^{\top} \bm{A}^{\top} \cdot \frac{\partial \bm{K}_{DD}}{\partial \bm{d_{ij}}} \right].
\end{array}
\]

The two partial derivatives of kernel matrices with respect to the current atom are sparse matrices with non-zero columns or rows only where their index is equal to the index of the current atom.
The two matrices are computed as follows:
\begin{equation}
\frac{\partial \bm{K}_{YD}}{\partial \bm{d}_{ij}} = \left[\begin{array}{ccccc}
0 & \cdots & \frac{\partial k\left(\bm{y}_{1}, \bm{d}_{j}\right)}{\partial \bm{d}_{ij}} & \cdots & 0 \\
\vdots & & \frac{\partial k\left(\bm{y}_{2}, \bm{d}_{j}\right)}{\partial \bm{d}_{ij}} & & \vdots \\
\vdots & & \vdots & & \vdots \\
0 & \cdots & \frac{\partial k\left(\bm{y}_{N} \bm{d}_{j}\right)}{\partial \bm{d}_{ij}} & \cdots & 0
\end{array}\right]
\label{eq:deriv1}
\end{equation}
and
\begin{equation}
\frac{\partial \bm{K}_{DD}}{\partial \bm{d}_{ij}} = \left[
\begin{array}{ccccc}
 0 & \cdots & \frac{\partial k\left(\bm{d}_{1}, \bm{d}_{j}\right)}{\partial \bm{d}_{ij}} & \cdots & 0 \\
 0 & \cdots & \frac{\partial k\left(\bm{d}_{2}, \bm{d}_{j}\right)}{\partial \bm{d}_{ij}} & \cdots & 0 \\
\vdots & & \vdots & & \vdots \\
\frac{\partial k\left(\bm{d}_{j}, \bm{d}_{1}\right)}{\partial \bm{d}_{ij}} & \cdots & \frac{\partial k\left(\bm{d}_{j}, \bm{d}_{j}\right)}{\partial \bm{d}_{ij}} & \cdots & \frac{\partial k\left(\bm{d}_{j}, \bm{d}_{m}\right)}{\partial \bm{d}_{ij}} \\
\vdots & & \vdots & & \vdots \\
 0 & \cdots & \frac{\partial k\left(\bm{d}_{n}, \bm{d}_{j}\right)}{\partial \bm{d}_{ij}} & \cdots & 0 \\
\end{array}\right].
\label{eq:deriv2}
\end{equation}

Depending on the kernel function of interest, the partial derivatives are computed via
\[
\frac{\partial k\left(\bm{x}, \bm{y}\right)}{\partial \bm{x}} = - \exp{\frac{- \|\bm{x}-\bm{y}\|^{2}_{2}}{2\sigma^2}}\frac{\left(\bm{x} - \bm{y}\right)}{\sigma^2}
\]
for the radial basis function kernel $k(\bm{x}, \bm{y}) = \exp(\frac{-\left\|\bm{x}-\bm{y}\right\|_{2}^{2}}{\sigma^{2}})$, and
\[
\frac{\partial k\left(\bm{x}, \bm{y}\right)}{\partial \bm{x}} = \beta \left( \bm{x}^{\top}\bm{y} + \alpha \right)^{\beta - 1} \bm{y}
\]
for the polynomial kernel $k(\bm{x}, \bm{y}) = (\bm{x}^{\top} \bm{y} + \alpha)^{\beta}$.

Since explicit solutions to the optimization problem on $\bm{d}_j$
do not seem to exist, we update each column $\bm{d}_{j}$ through gradient descent procedure applied on each element.
At gradient descent iteration $\ell$, the next element value is computed via
\[
\bm{d}_{ij}^{(\ell+1)} = \bm{d}_{ij}^{(\ell)} - \gamma \bm{g}_{ij}^{(\ell)},
\]
where $\gamma \in \mathbb{R}_{+}$ represents the chosen learning rate
and $\bm{g}_{ij}^{(\ell)}$ is the gradient
\[
\bm{g}_{ij}^{(\ell)} = \operatorname{Tr} \left[ \bm{A} \bm{Z} \left( \bm{Z}^{\top} \bm{A}^{\top} \cdot \frac{\partial \bm{K}_{DD}}{\partial \bm{d_{ij}}}  - 2 \cdot \frac{\partial \bm{K}_{YD}}{\partial \bm{d_{ij}}} \right) \right],
\]
where the matrices \eqref{eq:deriv1} and \eqref{eq:deriv2} are computed for the dictionary at iteration $\ell$.
The update step of the ORKDL-D algorithm is obtained by introducing
a few steps of gradient descent as described above for updating the dictionary $\bm{D}$, after the update of $\bm{A}$ described by Algorithm \ref{alg:RKDL}.

\subsection{Mixed Optimized Reduced Kernel Dictionary Learning}
The previous subsection presents an update procedure for the kernel vectors $\bm{D}$.
A possible drawback is that direct optimization of the objective of \eqref{eq:RKDL} may lead to a dictionary $\bm{D}$ with weaker 
representation power for the entire set of signals, $\bm{Y}$.
We present here a further improvement.
To keep the representation significance of $\bm{D}$, which is a natural
property to require in the context of kernel methods, we introduce
the standard DL objective \eqref{eq:DL} into the ORKDL-D problem, obtaining
\begin{equation}
\begin{array}{cl}
\displaystyle\min _{\boldsymbol{A}, \boldsymbol{Z}, \boldsymbol{D}, \boldsymbol{X}} & \|\varphi(\boldsymbol{Y})-\varphi(\boldsymbol{D}) \boldsymbol{A} \boldsymbol{Z}\|_{F}^{2} + \lambda \|\boldsymbol{Y}-\boldsymbol{D} \boldsymbol{X}\|_{F}^{2} \\
\text { s.t. } & \left\|\boldsymbol{z}_{\ell}\right\|_{0} \leq s_{z}, \ell=1: N \\
& \left\|\varphi(\boldsymbol{D}) \boldsymbol{a}_{j}\right\|=1, j=1: n \\
& \left\|\boldsymbol{x}_{\ell}\right\|_{0} \leq s_{x}, \ell=1: N \\
& \left\|\boldsymbol{d}_{j}\right\|=1, j=1: n,
\end{array}
\label{eq:MORKDL}
\end{equation}
where $\lambda \in \mathbb{R}_{+}$ is a penalty constant.
Since a mixed objective is proposed, we name this method
Mixed Optimized Reduced Kernel Dictionary Learning (MORKDL-D). 

By following two directions of optimization, we update the nonlinear representation while conserving the representation power of the linear space. The current optimization problem is solved similarly with the previous problems. The update of a column of $\bm{D}$ is slightly different. The gradient of the current atom uses also the partial derivative of the linear term and is
\[
\tilde{\bm{g}}_{ij} = \bm{g}_{ij} - \bm{e}_{ij} 
\]
where $\bm{e}_{ij}$ represents the $i$th element of the vector
$(\bm{Y} - \bm{D}\bm{X}) \cdot \bm{X}_{j}^{\top}$;
here, $\bm{X}_{j}^{\top}$ is the $j$th row of $\bm{X}$.
On the other hand, the linear sparse representation matrix $\bm{X}$ is computed
with the OMP algorithm.
As before, the nonlinear sparse representation matrix $\bm{Z}$ is computed using the Kernel OMP algorithm, since $\bm{Z}$ and $\bm{X}$ can be optimized independently when $\bm{A}$ and $\bm{D}$ are fixed.

\section{Experiments}
\label{sec:Experiments}

In this section we present the main results obtained with the proposed methods. We used three different data sets: Digits, MNIST and CIFAR-10. For each experiment we trained a kernel dictionary for representing the whole data set or a subset extracted by its corresponding label. For example, for the Digits data set we used all the signals during the training procedure, while for the MNIST and CIFAR-10 data sets we used signals specific to a selected label. For the MNIST data set we used label $5$, while for the CIFAR-10 data set we used the first label.

All the algorithms were implemented in Matlab and Python and were run on a Desktop PC with Ubuntu 20.04 as operating system. The PC has a processor with 16 cores, with a base frequency of 2.90 GHz (Max Turbo Frequency 4.80 GHz), and 80 GB RAM memory. During the experiments, we computed the objective function of \eqref{eq:RKDL} (we report the values $\|\varphi(\bm{Y})-\varphi(\bm{D})\bm{A}
\bm{Z}\|_{F} / \sqrt{mN}$ of the error per signal element) at each iteration and measured the overall execution time.
Note that for MORKDL-D we report the values of the same error, not of the objective of \eqref{eq:MORKDL}.
Of course, for KDL we compute the objective of \eqref{eq:KDL}, which corresponds to the standard case $\bm{D}= \bm{Y}$.
The error and execution time were computed by the mean on ten different rounds. As kernel function we used the radial basis function $k(\bm{x}, \bm{y}) = \exp(\frac{-\left|\bm{x}-\bm{y}\right|_{2}^{2}}{\sigma^{2}})$. The kernel parameters were chosen according to a grid search, and for all data sets we chose $\sigma = 10$. All the algorithms trained a kernel dictionary $\bm{A}$ of size $p = 20$, with sparsity level $s_{z} = 4$, using $10$ iterations.
For the reduced methods we initially trained a dictionary $\bm{D}$ of size $n = 50$, having a sparsity level of $s_{x} = 5$, with $10$ iterations of the AK-SVD algorithm. The dictionary $\bm{D}$ dimensions ensures a reduced kernel by having a smaller number of atoms compared to the number of data signals. For example, the Digits data set contains $N = 5000$ signals of size $m = 784$, while the MNIST data set consists of approximately $N = 6000$ signals of the same size, of the same class. From the CIFAR-10 data set we used $N = 5000$ signals, specific to the interest class, of size $m = 1024$.

In the ORKDL-D and MORKDL-D algorithms, the dictionary $\bm{D}$ is updated with three gradient descent iterations. For each of the three data sets we chose a learning rate that ensures a smooth decrease of the objective function.
We took $\gamma = 5 \cdot 10^{-4}$ for the Digits and MNIST data sets and
$\gamma = 6 \cdot 10^{-4}$ for the CIFAR-10 data set.
For all our experiments we used $\lambda = 1$.

\begin{figure}[!ht]
\centerline{\includegraphics[width=9.2cm]{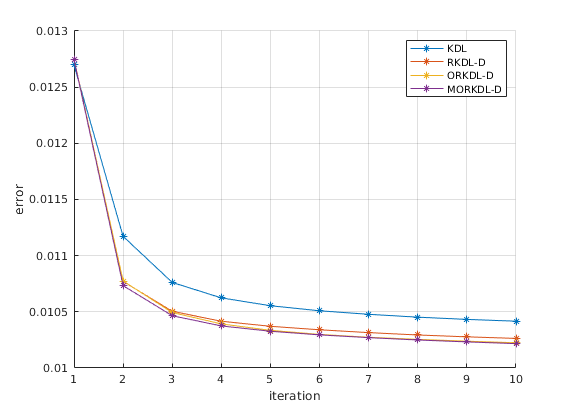}}
\caption{Digits - representation error per iteration}
\label{fig:Digits}
\end{figure}

\begin{figure}[!ht]
\centerline{\includegraphics[width=9.2cm]{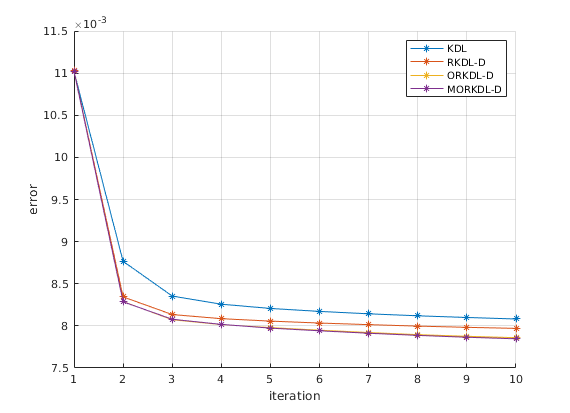}}
\caption{MNIST - representation error per iteration}
\label{fig:MNIST}
\end{figure}

\begin{figure}[!ht]
\centerline{\includegraphics[width=9.2cm]{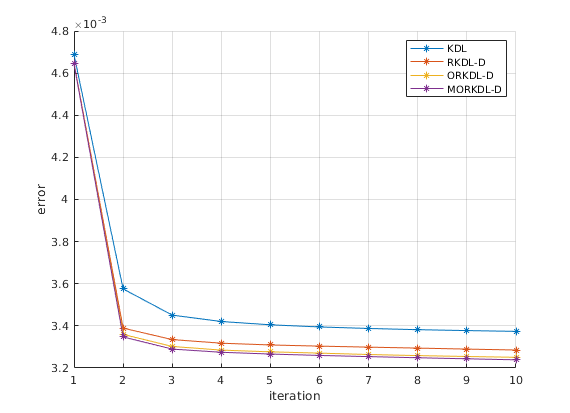}}
\caption{CIFAR-10 - representation error per iteration}
\label{fig:CIFAR10}
\end{figure}

For a better visualization of the results, we show the evolution of the nonlinear error (objective function of \eqref{eq:RKDL}) during the training procedure for each algorithm using the three data sets Digits (Figure \ref{fig:Digits}), MNIST (Figure \ref{fig:MNIST}) and CIFAR-10 (Figure \ref{fig:CIFAR10}).
As we can see, the improvements are visible from iteration two of training.
More results are given in two tables, containing the value of the error at the last iteration (Table \ref{tab:error}) and the execution time (Table \ref{tab:time}).
All the reduced methods achieve a smaller error compared to the KDL method. At first glance we notice that very large kernel spaces are not necessary to obtain good results. The introduction of a small dimensional space is enough to obtain satisfactory results.
For situations where an improvement of the nonlinear representation space is needed, several iterations of gradient descent can be run as needed. According to the results obtained on the three data sets it can be seen that the biggest advantage of the proposed methods is the reduction of the execution time. For example, the non-optimized reduced form obtains an execution time at least six times shorter than the KDL method. In the problems where we try to train the kernel space, an additional time will be introduced (for dictionary $\bm{D}$ update), but the execution times are still shorter than those of KDL. For reduced kernel methods the execution time is reduced by approximately 20-45\% with respect to the KDL problem. All the implementations are available at \href{https://github.com/denisilie94/rkdl}{https://github.com/denisilie94/rkdl}.

\begin{table}[!ht]
\caption{Last error value}
\centering
\begin{tabular}{|l||*{3}{c|}}\hline
\backslashbox{Algorithm}{Data set} & \makebox[5.5em]{Digits} 
& \makebox[5.5em]{MNIST} & \makebox[5.5em]{CIFAR-10} \\\hline\hline
KDL & 1.041 $\cdot 10^{-2}$ & 8.081 $\cdot 10^{-3}$ & 3.373 $\cdot 10^{-3}$ \\\hline
RKDL-D & 1.026 $\cdot 10^{-2}$ & 7.970 $\cdot 10^{-3}$ & 3.285 $\cdot 10^{-3}$ \\\hline
ORKDL-D & 1.022 $\cdot 10^{-2}$ & 7.859 $\cdot 10^{-3}$ & 3.250 $\cdot 10^{-3}$ \\\hline
MORKDL-D & 1.021 $\cdot 10^{-2}$ & 7.846 $\cdot 10^{-3}$ & 3.238 $\cdot 10^{-3}$ \\\hline
\end{tabular}
\label{tab:error}
\end{table}

\begin{table}[!ht]
\centering
\caption{Execution time in seconds}
\begin{tabular}{|l||*{3}{c|}}\hline
\backslashbox{Algorithm}{Data set} & \makebox[5.5em]{Digits} 
& \makebox[5.5em]{MNIST} & \makebox[5.5em]{CIFAR-10} \\\hline\hline
KDL & 61.7 & 92.5 & 65.1 \\\hline
RKDL-D & 9.2 & 10.9 & 9.6 \\\hline
ORKDL-D & 39.2 & 48.4 & 51.6 \\\hline
MORKDL-D & 38.9 & 49 & 51.4 \\\hline
\end{tabular}
\label{tab:time}
\end{table}

\section{Conclusions}
\label{sec:Conclusions}

In this paper we have presented a new approach to the Kernel Dictionary Learning problem by introducing a reduced kernel and thus obtaining a new algorithm, named RKDL.
This method is suitable for problems with large data sets, where standard KDL requires large memory sizes and long execution times.
Moreover, we have demonstrated that most of the time a large representation space for the kernel matrix is not always needed to obtain satisfactory results. The reduced form of the kernel is enough to get similar results or even better. Since the kernel matrix is smaller, the required execution time is also much shorter.

The RKDL algorithm was presented under three different forms: standard RKDL-D, optimized RKDL-D and mixed optimized RKDL-D, the latter two including optimization of the kernel vectors.
All of them obtain competitive results with the standard KDL problem.

\bibliographystyle{unsrt}
\bibliography{bib}

\end{document}